 \documentclass{ws-ijmpd}

\usepackage{graphicx}
\usepackage{epstopdf}
\usepackage{epsfig}
 \newcommand{\be}[1]{\begin{equation}\label{#1}}
 \newcommand{\ee}{\end{equation}}

\begin{document}

\markboth{Shuang-Nan Zhang} {``Frozen Star" Paradox, Astrophysical Black Holes, Singularity \& Birkhoff's
Theorem}

%
\catchline{}{}{}{}{}
%

\title{On the Solution to the ``Frozen Star" Paradox, Nature of Astrophysical Black Holes,
non-Existence of Gravitational Singularity in the Physical Universe and Applicability of the Birkhoff's Theorem}

\author{Shuang-Nan Zhang}

\address{Key Laboratory of Particle Astrophysics\\
Institute of High Energy Physics\\
Chinese Academy of Sciences\\
Beijing, 100049, China\\
zhangsn@ihep.ac.cn}

\maketitle

\begin{history}
\received{Day Month Year}
\revised{Day Month Year}
\comby{Managing Editor}
\end{history}

\begin{abstract}
Oppenheimer and Snyder found in 1939 that gravitational collapse in vacuum produces a ``frozen star", i.e., the
collapsing matter only asymptotically approaches the gravitational radius (event horizon) of the mass, but never
crosses it within a finite time for an external observer. Based upon our recent publication on the problem of
gravitational collapse in the physical universe for an external observer, the following results are reported
here: (1) Matter can indeed fall across the event horizon within a finite time and thus BHs, rather than
``frozen stars", are formed in gravitational collapse in the physical universe. (2) Matter fallen into an
astrophysical black hole can never arrive at the exact center; the exact interior distribution of matter depends
upon the history of the collapse process. Therefore gravitational singularity does not exist in the physical
universe. (3) The metric at any radius is determined by the global distribution of matter, i.e., not only by the
matter inside the given radius, even in a spherically symmetric and pressureless gravitational system. This is
qualitatively different from the Newtonian gravity and the common (mis)understanding of the Birkhoff's Theorem.
This result does not contract the ``Lemaitre-Tolman-Bondi" solution for an external observer.
\end{abstract}

\keywords{Frozen Star; Black Hole; Singularity; Birkhoff's Theorem; Shapiro Delay.}

\section{Introduction}

Oppenheimer \& Snyder\cite{OS39} studied the problem of black hole (BH) formation from gravitational collapse
and arrived at two conclusions which have deeply influenced our understanding of astrophysical BH formation ever
since. The first conclusion is: ``The total time of collapse for an observer comoving with the stellar matter is
finite." However it should be realized that the comoving observer is also within the event horizon with the
collapsing matter, once a BH is formed. The second and last conclusion of the paper is: ``An external observer
[{\bf O} hereafter] sees the star asymptotically shrinking to its gravitational radius [the radius of the event
horizon of the BH of the same mass, $R_{\rm H}=2GM/c^2$ hereafter.]." This means that {\bf O} will never witness
the formation of an astrophysical BH. Given the finite age of the universe and the fact that all observers are
necessarily external, the last conclusion of Ref.~\refcite{OS39} seems to indicate that astrophysical BHs cannot
be formed in the physical universe through gravitational collapse.

Recently, Vachaspati, Stojkovic \& Krauss\cite{Krauss} have stressed that ``The process of BH formation is
generally discussed from the viewpoint of an infalling observer. However, in all physical settings it is the
viewpoint of the asymptotic observer [i.e. {\bf O}] that is relevant." They analyzed the process of the
self-collapse of a domain wall (a massive shell with no thickness) and concluded that {\bf O} sees the domain
wall asymptotically shrinking to $R_{\rm H}$, i.e., a BH is never formed within a finite time to {\bf O}. This
is a further confirmation to the conclusion of Ref.~\refcite{OS39}. Vachaspati et al.\cite{Krauss} then went on
to study the quantum mechanical effect of the contracting shell and found that the matter accumulating just
outside $R_{\rm H}$ actually produces radiation, which they called pre-Hawking radiation. They concluded that
``Evaporation by pre-Hawking radiation implies that {\bf O} can never lose objects down a BH."

Combining the above two works separated by nearly 70 years, a very surprising scenario seems inevitable:
Gravitational collapse will not produce BHs, but result in complete conversion of matter into radiation. This
scenario, if correct, would have profound implications to our understanding of general relativity which has long
been considered to robustly predict the existence of BHs, as well as a vast amount of astronomical observations
which can, and perhaps only, be understood by invoking BHs\cite{Zhang}. However, both of the above works are
over-simplified and do not catch all the essence of gravitational collapse in the physical universe, because
both investigations only considered gravitational contraction in vacuum and the work of Ref.~\refcite{Krauss}
did not allow a finite thickness of the contracting shell.

To overcome the drawbacks of these two works discussed above, Liu \& Zhang\cite{Liu} studied the gravitational
collapse of a single shell and double-shells onto a pre-existing BH; these shells can have finite thicknesses
and the outer shell in the double-shell case mimics the matter outside the collapsing shell in the physical
universe. The gravitational contractions studied in the two previous works can be considered as special cases of
that studied in Ref.~\refcite{Liu}. The main conclusion of Liu \& Zhang\cite{Liu} is that matter does not
accumulate outside $R_{\rm H}$, but instead falls straight across it, within a finite time of {\bf O}. In the
rest of this paper, we first review briefly the main results in Ref.~\refcite{Liu}, and then discuss several
issues related to the ``frozen star" paradox, nature of astrophysical BHs, gravitational singularity in the
physical universe, and finally applicability of the Birkhoff's theorem. All calculations and discussions in this
report are within the framework of Einstein's general relativity.

\section{Exact solutions for shells collapsing onto a pre-existing BH}

We briefly review the main results in Ref.~\refcite{Liu}. In Fig.~1, we show the initial conditions of
gravitational collapse onto a BH in the comoving coordinates, for a single shell and double-shell cases,
respectively. For the single shell case, $m$ and $m_s $ are the total gravitating masses of the BH and the
shell, respectively. $a'$ and  $a$ are the radii of the inner and outer boundaries of the shell, respectively.
Letting $a'=0$ and $m=0$, this recovers to the case studied in Ref.~\refcite{OS39}. Letting $a\approx a'$ and
$m=0$, this recovers to the case studied in Ref.~\refcite{Krauss}.

\begin{figure}[pb]
\hbox{\epsfig{file=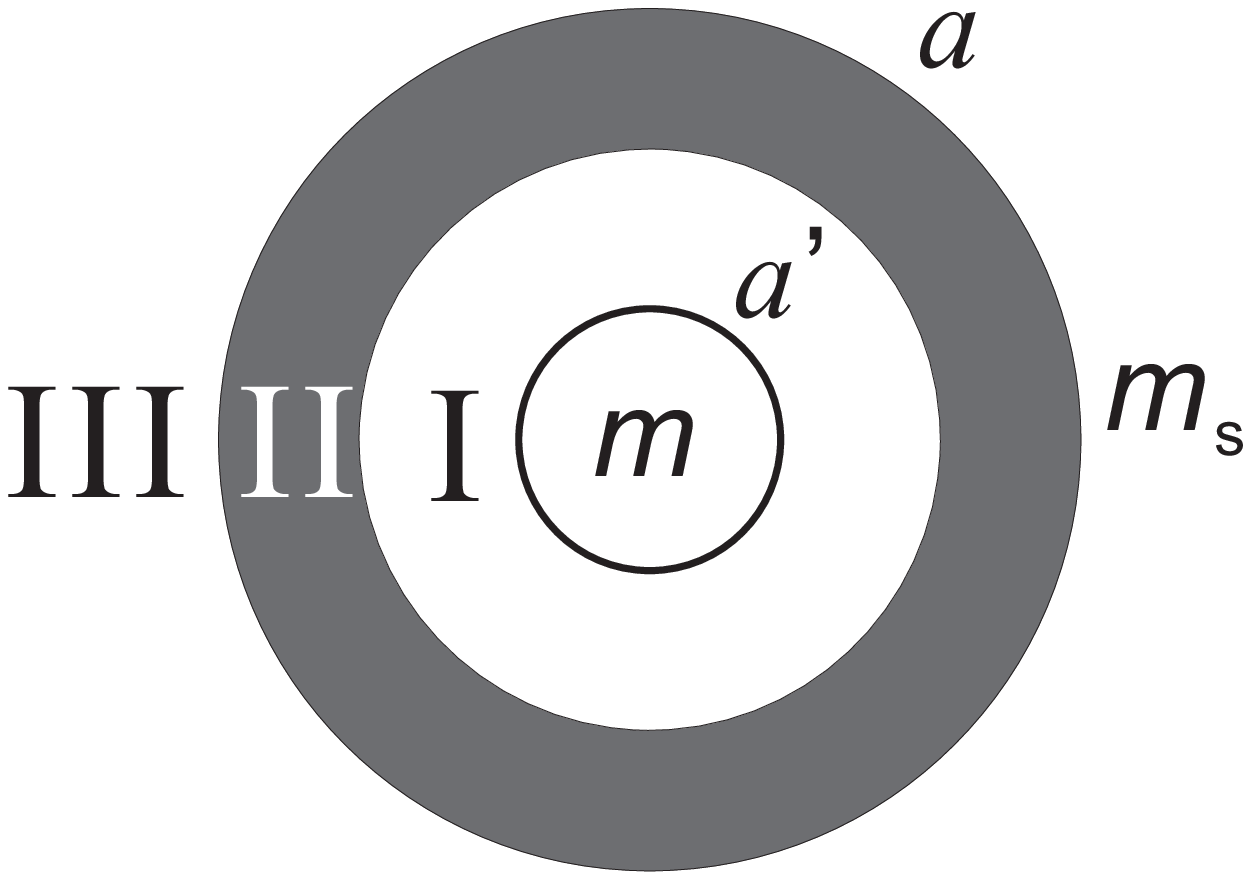,width=6cm}\epsfig{file=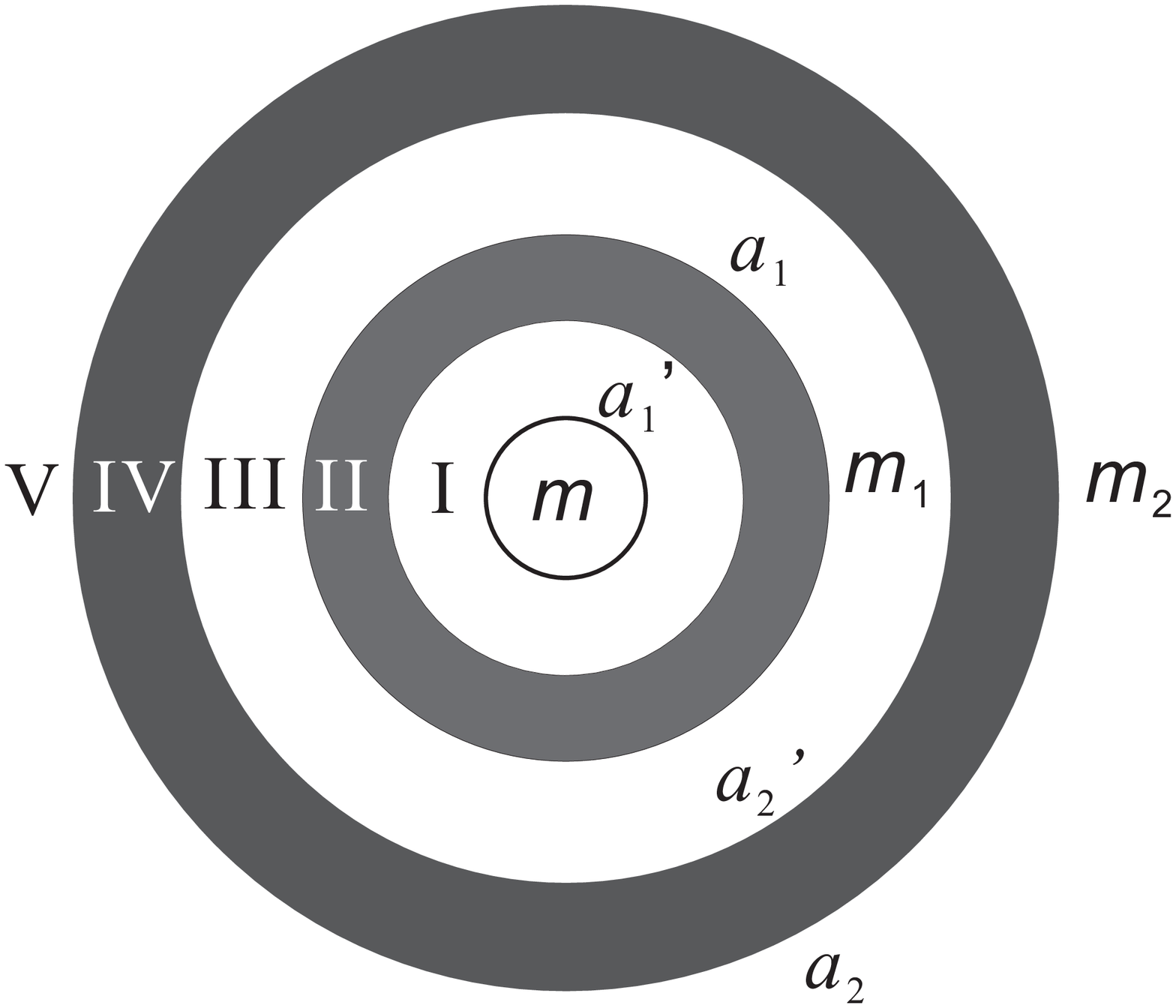,width=6.7cm}} \caption{Initial conditions of
gravitational collapse onto a BH in the comoving coordinates. {\it Left panel}: the case for one shell. $m$ and
$m_s $ are the total gravitating masses of the BH and the shell, respectively. $a'$ and  $a$ are the radii of
the inner and outer boundaries of the shell, respectively. {\it Right panel}: the case for two shells.  $m$,
$m_1 $ and $m_2 $ are the total gravitating masses of the BH, shells 1 and 2, respectively. $a_1 '$ and $a_1$
are the radii of the inner and outer boundaries of shell 1, respectively.  $a_2 '$ and  $a_2$ are the radii of
the inner and outer boundaries of shell 2, respectively. \label{f1}}
\end{figure}

We first solve the field equations in the comoving coordinates, following Ref.~\refcite{OS39}; the metric or
extrinsic curvature is required to be continuous across the boundaries of the three regions. We then transform
the solutions in the three regions to the Schwarzschild coordinates (for {\bf O}). In each of the three regions
shown in the left panel of Fig.~1, the metric can be expressed in the Schwarzschild-like form, \be{eq1} ds^2  =
h_i (1-\frac{2M(r)}{r})dt^2  - (1-\frac{2M(r)}{r})^{-1}dr^2  - r^2 d\Omega^2, \ee where $M(r)$ is the total
gravitational mass within $r$, and $i=1$, 2, or 3, specifies region I, II, or III, respectively. For region III,
obviously $h_3=1$, i.e., the metric is exactly Schwarzschild. In region II, $h_2=h_2(t,r)<1$, though its
analytic form cannot be obtained generally. In region I, the continuity condition ensures that $h_1=h_1(t)<1$.
Therefore the metric in either region I or II is not Schwarzschild, because both $h_1$ and $h_2$ change as the
shell falls in.

\begin{figure}[pb]
\hbox{\epsfig{file=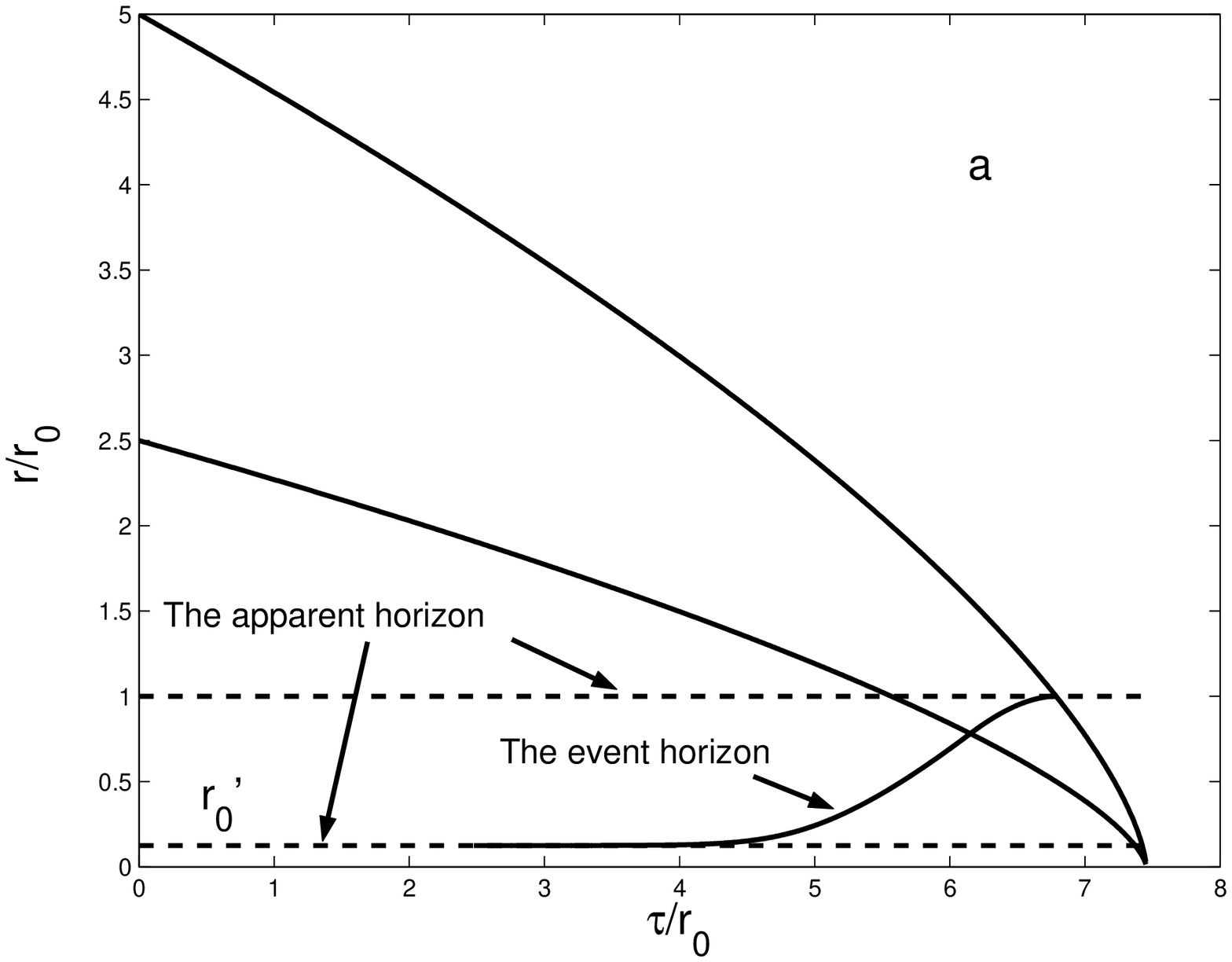,width=6.3cm}\epsfig{file=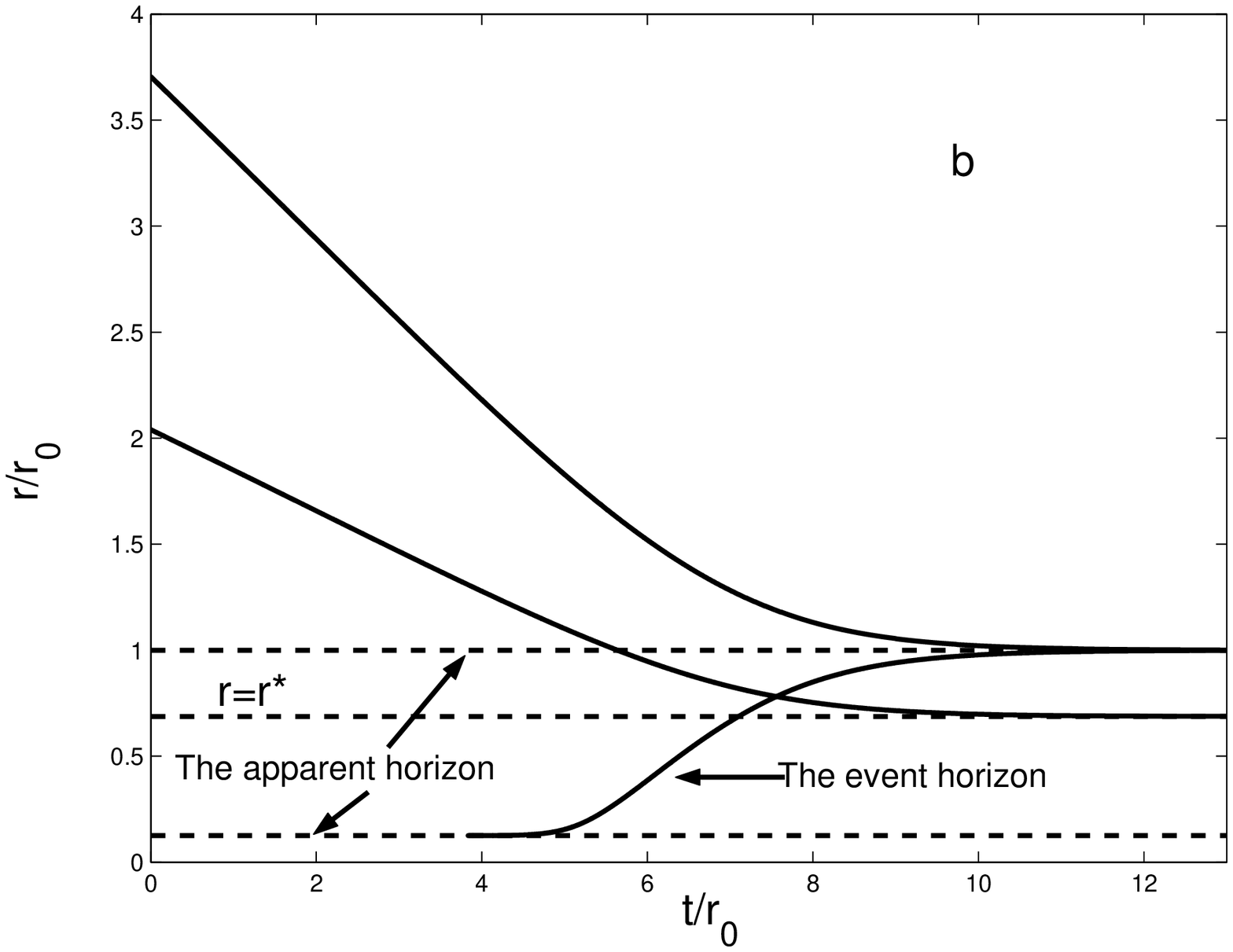,width=6.3cm}} \caption{The solution for the one
shell case. (a) and (b) are evolution curves for  $a = 5r_0 $,  $a' = 2.5r_0 $, and $r_0 ' = 1/8r_0$ with
comoving time and coordinate time, respectively. The evolution of the event and apparent horizons are also
shown. Here $G=c=1$, $r_0 '=2m$ and $r_0=2(m+m_s)$. \label{f2}}
\end{figure}

The motion of the shell is shown in Fig.~2, in both the comoving and external coordinates. Clearly the shell
falls into the BH within a finite comoving time. For {\bf O}, the body of the shell also crosses $R_{\rm H}$
within a finite time, except for its outer boundary which asymptotically approaches to $R_{\rm H}$. This result
is consistent with that of Ref.~\refcite{OS39}, but qualitatively different from that of Ref.~\refcite{Krauss}.
The difference is due to the finite thickness of the shell in Fig.~1, in contrast to the domain wall assumption
in Ref.~\refcite{Krauss}. In the case of a finite thickness of the shell, the increase of $R_{\rm H}$ swallows
the shell; obviously this cannot happen if the shell has no thickness. We therefore dismiss the conclusion of
Ref.~\refcite{Krauss}, because there is no matter available outside the event horizon to produce any radiation,
for a physical shell with non-zero thickness.

\begin{figure}[pb]
\hbox{\epsfig{file=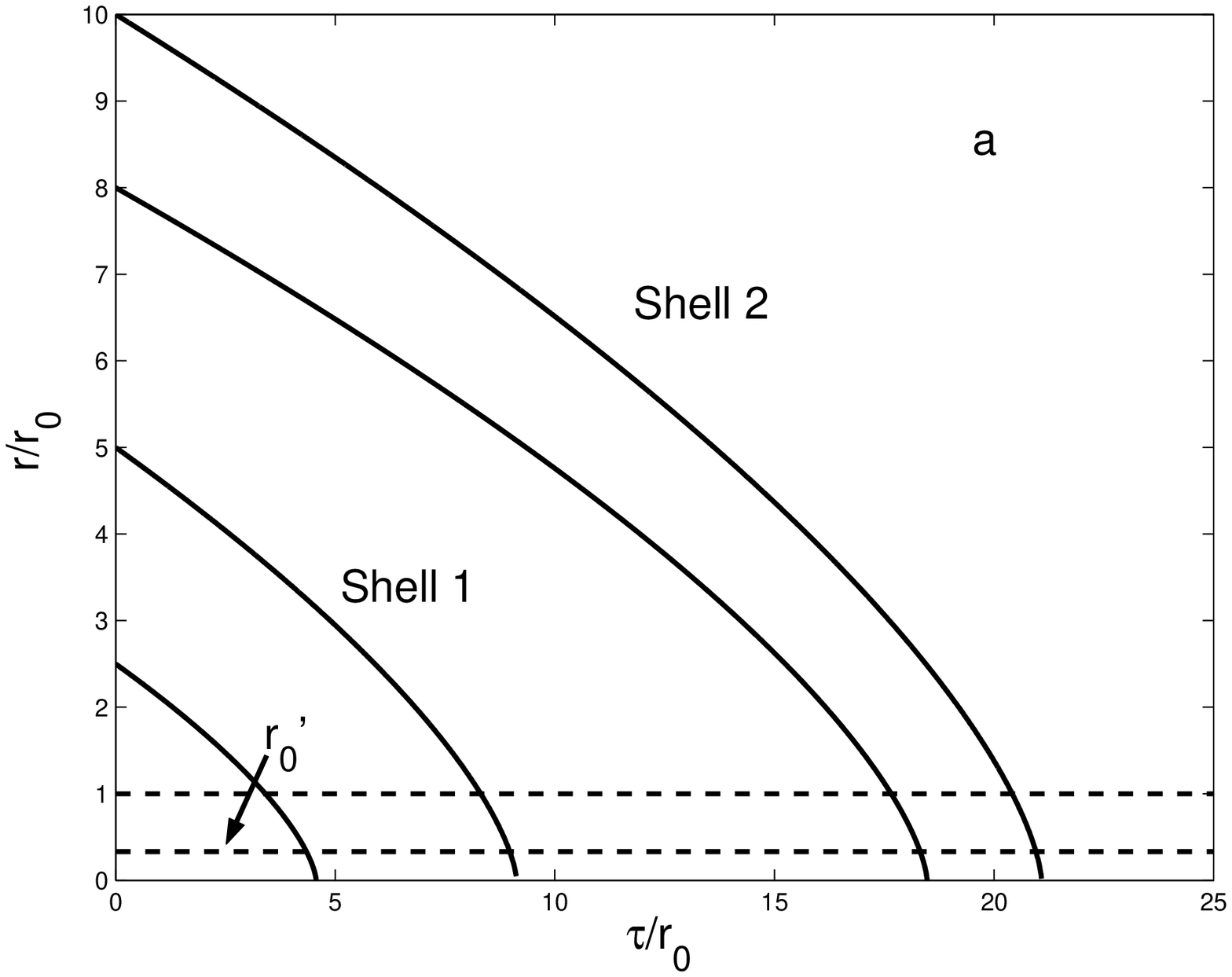,width=6.3cm}\epsfig{file=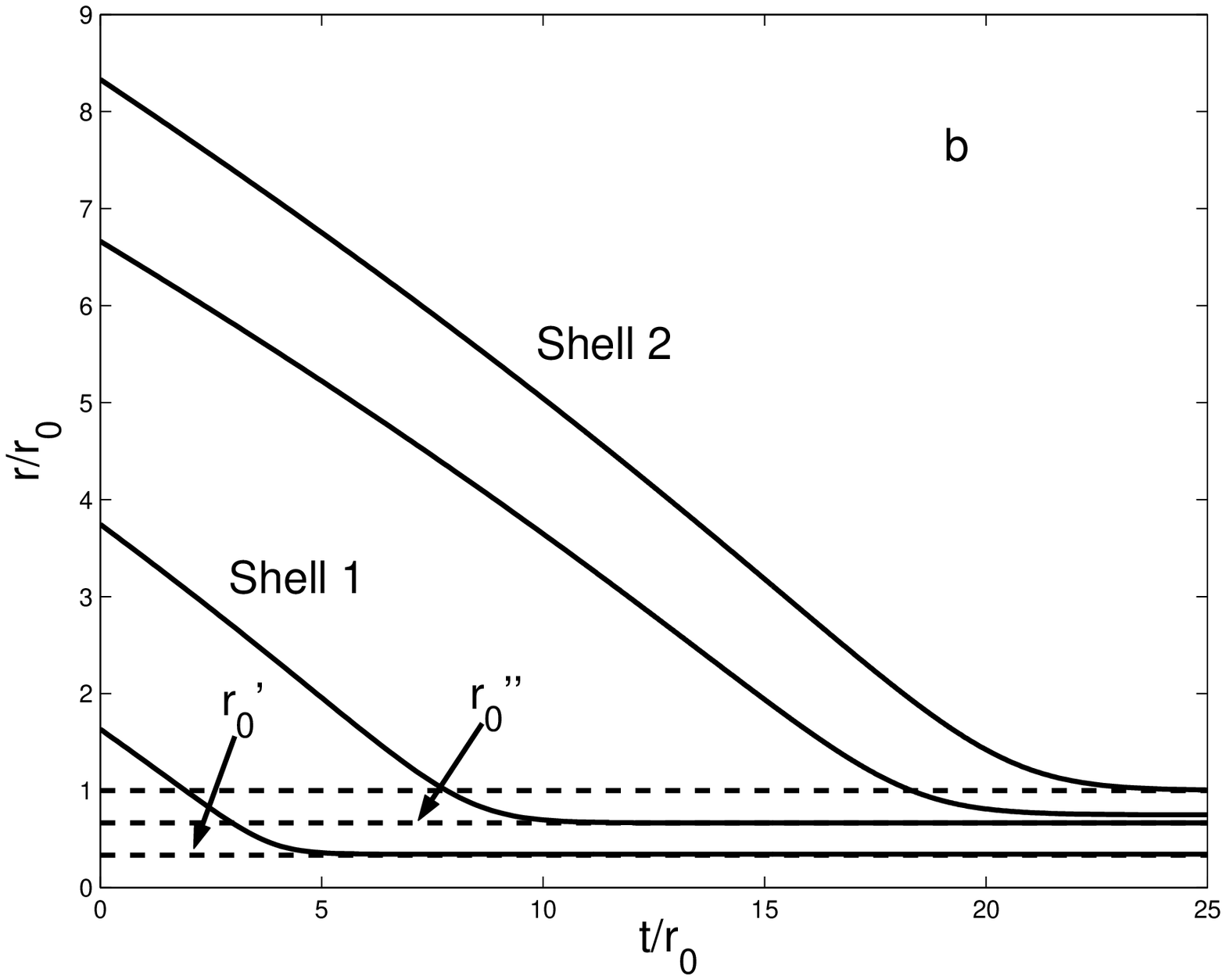,width=6.3cm}} \caption{The solution for the
double-shell case. (a) and (b) are evolution curves for  $a_2  = 10r_0 $, $a_2 ' = 8r_0 $,  $a_1 = 5r_0 $, $a_1
' = 2.5r_0 $, $r_0 ' = 1/3r_0 $, and $r_0 '' = 2/3r_0 $ with comoving time and coordinate time, respectively.
Here $G=c=1$, $r_0 '=2m$, $r_0 ''=2(m+m_1)$, and  $r_0 =2(m+m_1+m_2)$.\label{f3}}
\end{figure}

However the calculations shown in Fig.~2 still neglects one important fact for gravitational collapse in the
physical universe. There is always some additional matter between the observer and the infalling shell being
observed (we call it shell 1 Fig.~1 (right)), and the additional matter (we call it shell 2 in Fig.~1 (right))
is also attracted to fall inwards by the inner shell and the BH. We thus calculate the motion of the
double-shell system. For {\bf O}, the metric in each of the five regions still takes the form of Eq.~(1), with
$h=1$ in region V, but $h<1$ and is also time-dependent in all other four regions. The motions of both shells
are shown in Fig.~3. In this case, shell 1 can cross $R_{\rm H}$ completely even for {\bf O}.

\section{``Frozen Star" or Black Hole?}

The asymptotic behavior of the gravitational collapse is related to a well-known novel phenomenon predicted by
general relativity, i.e., {\bf O} sees a test particle falling towards a BH moving slower and slower, becoming
darker and darker, and is eventually frozen near the event horizon of the BH. This process was also vividly
described and presented in many popular science writings\cite{Ruffini,Luminet,Thorne,Begelman} and
textbooks\cite{Misner,Weinberg,Shapiro,Schutz,Townsend,Raine}. Because of this, the object of a complete
gravitational collapse has been called a ``frozen star". A fundamental question can then be asked: ``Does a
gravitational collapse form a frozen star or a BH?" Alternatively one can also ask: ``Can any matter ever fall
into a BH, even if it does exist?" In both questions, the clock of {\bf O} is referred to.

The answers to the above questions have been debated for decades. One answer is that since the comoving observer
indeed has observed the test particle falling through the event horizon and reaching the singularity point, then
in reality matter indeed has fallen into the BH and reached the singularity point. However, since {\bf O} has no
way to communicate with the comoving observer once matter crosses $R_{\rm H}$, {\bf O} has no way to `know' if
the test particle has fallen into the BH. The other answer is to invoke quantum effects. It has been argued that
quantum effects may eventually bring matter into a BH, as seen by {\bf O}\cite{Frolov}. However, as pointed out
recently\cite{Krauss}, even in that case the BH will still take an infinite time to form and the pre-Hawking
radiation will be generated by the accumulated matter just outside the event horizon. Thus both answers fail in
the real world.

In desperation, we may take the attitude of ``who cares?" When the test particle is sufficiently close to
$R_{\rm H}$, the redshift is so large that practically almost no signals from the test particle can be seen by
{\bf O} and apparently the test particle has no way of turning back, therefore the ``frozen star" does appear
``black" and is an infinitely deep ``hole". For practical purposes we may still call it a ``BH", whose total
mass is also increased by the infalling matter. Apparently this is the view taken by most people in the
astrophysical community; this is demonstrated by those similar arguments in many well-known
textbooks\cite{Misner,Weinberg,Shapiro,Schutz,Townsend,Raine,Hawking} and popular science writings by many
well-known scientists\cite{Ruffini,Luminet,Thorne,Begelman}. This is the reason that the ``frozen star"
terminology has almost disappeared completely from professional literature, although the issue had not been
fully understood until very recently.

For example, recently Vachaspati et al.\cite{Krauss} pointed out that matter accumulating just outside $R_{\rm
H}$ would produce pre-Hawking radiation. More than that, when two such ``frozen stars" merge together,
electromagnetic radiations will be released, in sharp contrast to the merging of two genuine BHs [i.e., all
their masses are within $R_{\rm H}$]; the latter can only produce gravitational wave radiation\cite{Vachaspati}.
Therefore the physical properties of ``frozen stars" are fundamentally different from BHs. We therefore must
answer these questions definitively.

Finally these questions are answered definitely and the above ``frozen star" paradox is solved completely by Liu
\& Zhang\cite{Liu}. As shown in Figs.~2 and 3 taken from Ref.~\refcite{Liu}, matter cannot accumulate outside
$R_{\rm H}$, due to the increase of $R_{\rm H}$ which swallows the matter falling in. The fundamental reason for
the asymptotic behavior of a test particle is due to the negligence of the influence of the test particle to the
global properties of the whole gravitating system, therefore $R_{\rm H}$ would not change during the infalling
process of the test particle. Therefore a BH can indeed be formed from gravitational collapse, and ``frozen
stars" cannot exist in the physical universe.

\section{Black Hole or Singularity?}

A BH has always been considered as a spacetime singularity. However Zhang\cite{Zhang} classified BHs into three
classes: mathematical BHs, physical BHs or astrophysical BHs. A mathematical BH is the vacuum solution of
Einstein's field equations of a point-like object, whose mass is completely concentrated at the center of the
object, i.e., the singularity point. A physical BH is an object whose mass and charge are all within $R_{\rm
H}$, regardless of the distribution of matter within; consequently a physical BH is not necessarily a
mathematical BH. Finally an astrophysical BH is a physical BH, which can be formed through astrophysical
processes in the physical universe and within a time much shorter than or at most equal to the age of the
universe.

From Figs.~2 and 3, it is clear that matter can never arrive at the singularity point, according to the clock of
{\bf O}. This means that astrophysical BHs in the physical universe are not mathematical BHs. Given that we do
not yet know for sure if there are other channels (other than through gravitational collapse of matter) of
forming BHs in the physical universe, we therefore suggest that spacetime singularity does not exist. This
conclusion may sound surprising and against the common understanding of general relativity and BH physics.
However we do not seem to have other alternatives, because we can only observe and study the formation process
of an astrophysical BH from outside $R_{\rm H}$, and thus for us, as external observers, matter can never arrive
at the singularity point even after crossing $R_{\rm H}$ and loss communications from us.

\section{Applicability of the Birkhoff's Theorem}
\begin{figure}[pb]
\hbox{\epsfig{file=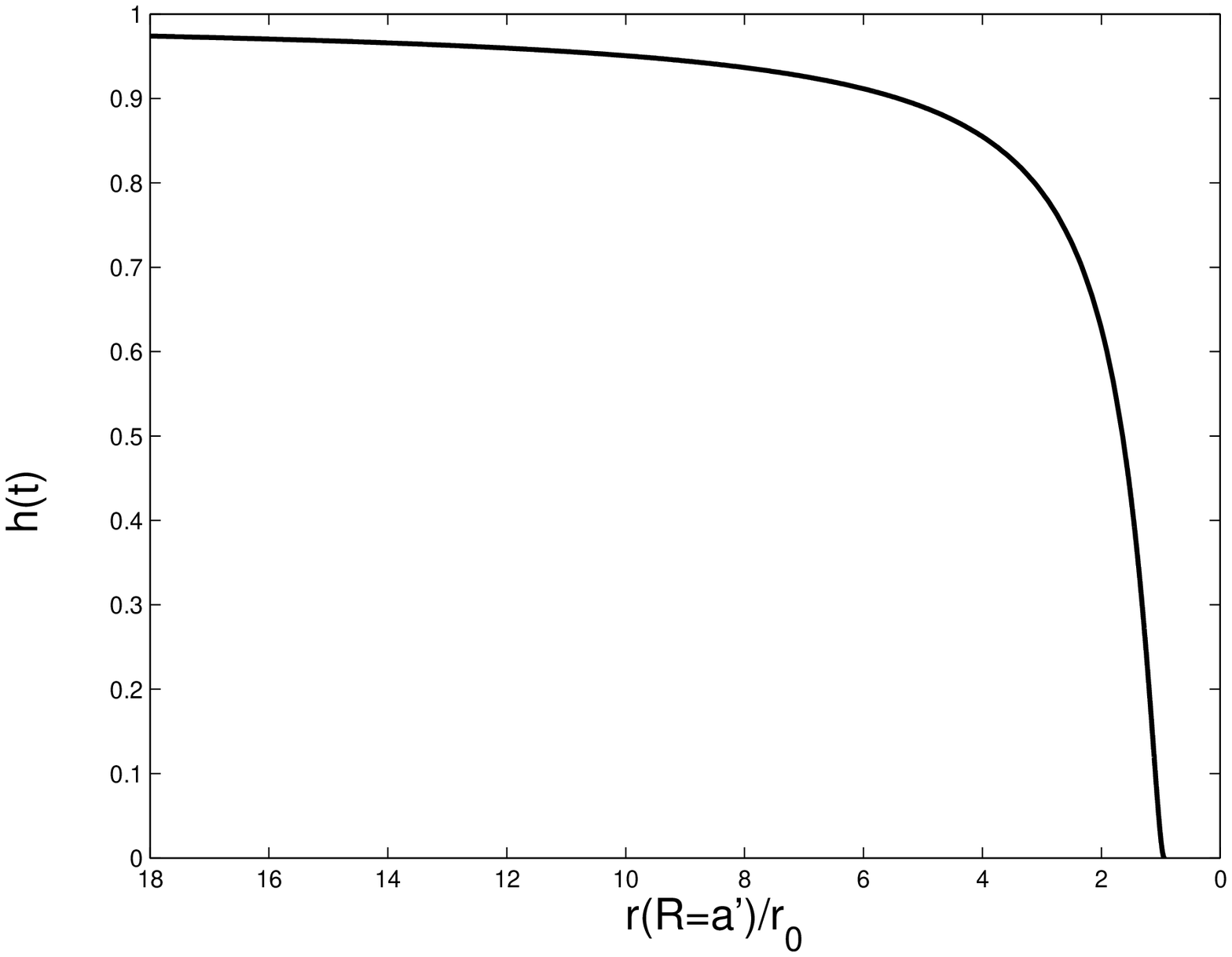,width=6.3cm}\epsfig{file=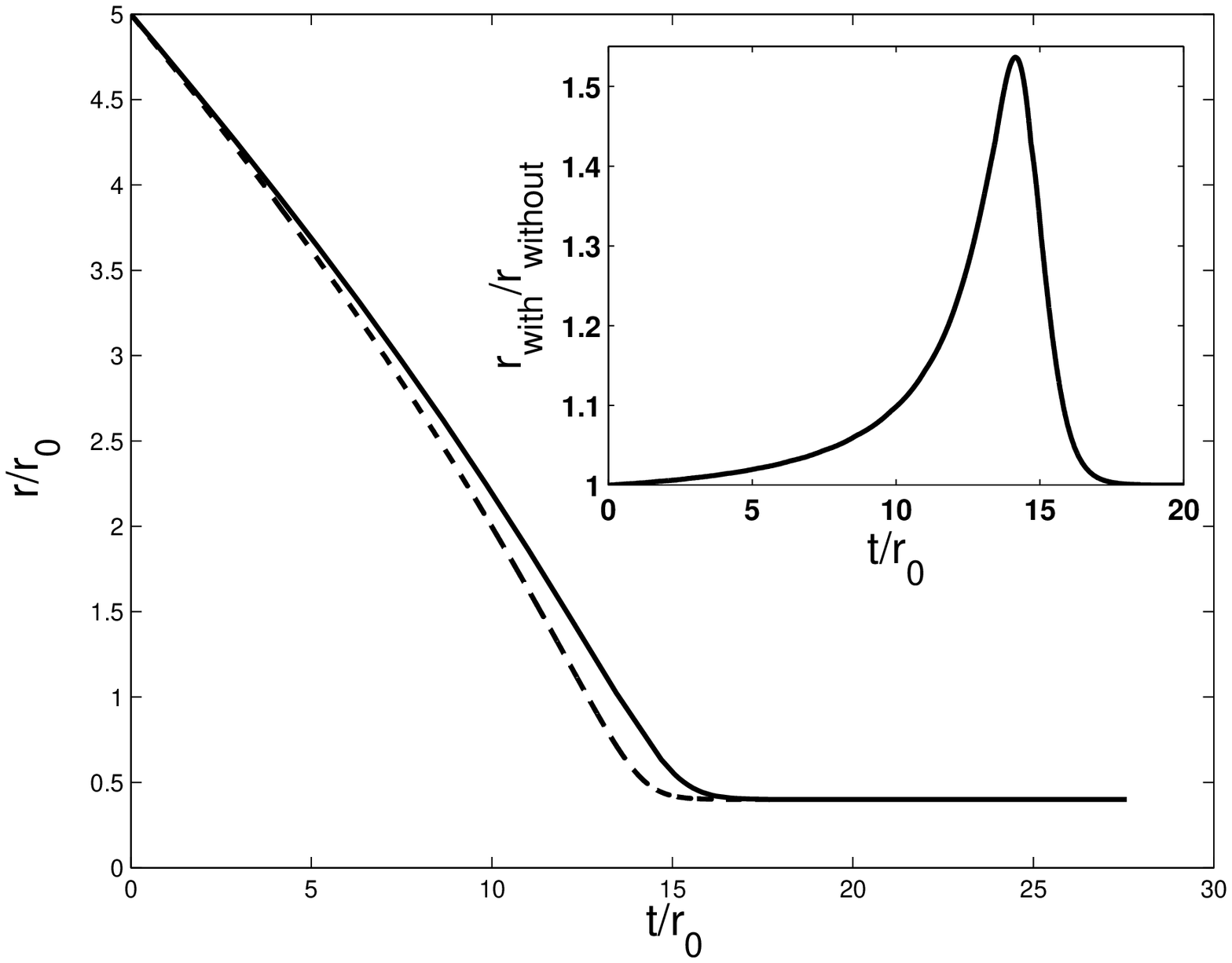,width=6.3cm}} \caption{{\it Left panel:} the
evolution of $h(t)$ with the position of the outer boundary of the shell for the one shell case with $a = 5r_0
$,  $a' = 2.5r_0 $, and $r_0 ' = 1/8r_0 $. {\it Right panel:} the comparison of the evolution of the outer
boundary of shell 1 between the case with (solid) or without (dashed) shell 2. The parameters of the shells are
$a_2 = 10r_0 $, $a_2 ' = 6r_0 $, $a_1 ' = 2.5r_0 $, $a_1  = 5r_0 $, $r_0 ' = 1/5r_0 $, and  $r_0 '' = 2/5r_0 $,
where $r_0$ is the Schwarzschild radius corresponding to the total gravitating mass of the system. The inset is
the ratio of the solid line to the dashed line. \label{f5}}
\end{figure}

The Birkhoff's theorem states that the metric in the {\it vacuum} for a spherically distributed gravitational
system is static and Schwarzschild. This means that only the {\ exterior} metric of a spherically distributed
gravitational system is Schwarzschild, e.g., the metric in region I of Fig.~1 (left) and regions I \& III of
Fig.~1 (right) is not necessarily Schwarzschild, although there is no matter in these regions. Actually the
proof of the Birkhoff's Theorem requires that there is no matter between the given location to infinity;
otherwise the time coordinate cannot be taken in the same way as that in the Schwarzschild metric. This is
generally not well appreciated by researchers. For example, in doing gravitational lensing calculations, the
metric everywhere is always taken as Schwarzschild, i.e., $h=1$ in Eq.~1, if the system is spherically
symmetric. This is obviously incorrect, as $h<1$ except in the exterior region, i.e., the real vacuum. If, on
the other hand, one forces $h=1$ even in region I of Fig.~1 (left), the metric across the boundary between
regions I and II would be discontinuous, i.e., non-physical.

Because Shapiro delay normally refers to the case $h=1$ and $h<1$ means extra delay, we call the time delay in
the case of $h<1$ {\it Generalized} Shapiro delay, which should be considered in calculating the light
propagation time through, e.g., the dark matter halos of galaxies or clusters of galaxies. This means that
negligence of this extra delay would over-estimate the mass of the system under investigation.

Fig.~4 (left) shows that $h<1$ and is a function of time, i.e., the metric in region I is neither schwarzschild
nor stationary, although there is no matter there. Fig.~4 (right) further shows that the motion of shell 1 (the
inner shell) is influenced by the existence and motion of shell 2 (the outer shell). This is clearly against the
common misconception that metric is only determined by the interior mass; this is a fundamental difference
between Einstein's general relativity and Newtonian gravity. However, this conclusion seems to contradict the
well-known ``Lemaitre-Tolman-Bondi" metric inside a spherically symmetric distribution of matter, \be{eq2} ds^2
= dT^2-\frac{(\sqrt{E+\frac{2M}{r}}dT+dr)^2}{1+E}-r^2d\Omega^2, \ee which is expressed in the Painl$\acute{{\rm
e}}$v-Gullstrand coordinates\cite{Lasky}, and where $E=E(T, r)$ is the energy function of the shell at $r$ and
$M=M(T, r)$ is the gravitational mass inside $r$. Clearly the metric at $r$ is fully determined by $E(T, r)$ and
$M(T, r)$.

To understand this apparent conflict, it is necessary to transform $T$ in the Painl$\acute{{\rm e}}$v-Gullstrand
coordinates to $t$ in the Schwarzschild coordinates by\cite{Lasky}, \be{eq3} (\frac{\partial T}{\partial
t})^2=1+E\ \ \ {\rm and} \ \ \ (1-\frac{2M}{r})\frac{\partial T}{\partial r}=\sqrt{\frac{2M}{r}+E}. \ee Solving
these partial differential equations requires integrals from $r$ to infinity, or to the outer boundary of the
system to match the Schwarzschild metric. Therefore eventually the metric at $r$ includes both $E$ and $M$
outside $r$, if one uses the clock of {\bf O}. For example, $h(r)$ in Eq.~1 has to be calculated inside the
system when tracing light through it. Therefore, taking $h(r)=1$, as commonly done, is only an approximation.

\section*{Acknowledgments}

The author thanks Profs. Remo Ruffini of ICRANet for discussion on the applicability of the Birkhoff's theorem,
Jialu Zhang of USTC for comment on the ``frozen star" paradox, and Pengjie Zhang of SHAO for discussion on the
``Lemaitre-Tolman-Bondi" solution. The research presented here is partially supported by the National Natural
Science Foundation of China under grant Nos. 10821061, 10733010, 0725313, and by 973 Program of China under
grant No. 2009CB824800.

\end{document}